\documentstyle[11pt]{article}

\textwidth\def\kon#1#2{\vbox{\halign{##&&##\cr\lower4pt
\hbox{$\scriptsrcriptstyle\vert$}\hrulefill &\hrulefill\lower4pt
\hbox{$\scriptscriptstyle\vert$}\cr $#1$&$#2$\cr}}}

\def\d{\partial}
\def\=d{\,{\buildrel\rm def\over =}\,}

\def\sqr#1#2{{\vcenter{\vbox{\hrule height.#2pt\hbox{\vrule width.
#2pt height#1pt \kern#1pt \vrule width.#2pt}\hrule height.#2pt}}}}
 
\def\eps{\varepsilon}
\def\curl{{\rm curl}}

\def\eh{{\scriptstyle{1\over 2}}}

\def\lap{\bigtriangleup}
\def\rho{\varrho}

\begin{document}
\baselineskip=20pt
\title{Electrophysiology of living organs from first principles}
\author{G\"unter Scharf
\footnote{e-mail: scharf@physik.unizh.ch}
\\ Institut f\"ur Theoretische Physik, 
\\ Universit\"at Z\"urich, 
\\ Winterthurerstr. 190 , CH-8057 Z\"urich, Switzerland
\\
\\Lam Dang and Christoph Scharf
\footnote{e-mail: christoph.scharf@gmail.com}
\\Cardiac Electrophysiology, Clinic im Park,
\\Seestrasse 220, CH-8027 Z\"urich, Switzerland}

\date{}

\maketitle\vskip 3cm

\begin{abstract}
Based on the derivation of the macroscopic Maxwell's equations
by spatial averaging of the microscopic equations, we discuss 
the electrophysiology of living organs. Other methods of averaging
(or homogenization) like the bidomain model are not compatible with 
Maxwell's theory.
We also point out that modelling the active cells by source currents
is not a suitable description of the situation from first principles.
Instead, it turns out that the main source of the measured electrical
potentials is the polarization charge density which exists
at the membranes of active cells and adds up to a macroscopic polarization.
The latter is the source term in the Laplace equation, the solution of
which gives the measured far-field potential. As a consequence it is
the polarization or dipole density which is best suited for localization
of cardiac arrhythmia.

\end{abstract}
\vskip 1cm
{\bf keywords:} bioelectric sources, electrocardiography, 
cardiac action potential
\newpage

\section{Introduction}

Macroscopic measurements of the electrical activity of living organs
are frequently employed to diagnose diseases. The main examples of
bioelectric recordings are electrocardiography of the heart (ECG),
electroencephalography of the brain (EEG) and electromyography 
of the muscles (EMG). The purpose of
the present article is to describe the origin of the measured electrical
potentials in these techniques. The basis of this description
are the macroscopic Maxwell's equations. Consequently, the physically
well defined quantities appearing in these equations are the primary
quantities to be used, and it is highly desirable to relate the measured
potentials to them. Such a model-independent description of the electrical
activity of the organ without ad hoc model assumptions (see below)
has great clinical value, for example for the localization of cardiac
arrhythmia as discussed at the end of sect.3.

In electrodynamics one has to make a clear distinction between
Maxwell's equation in the vacuum
$${\d\vec B\over\d t}=-\nabla\times\vec E\eqno(1.1)$$
$$\nabla\cdot\vec B=0\eqno(1.2)$$
$$\eps_0{\d\vec E\over\d t}={1\over\mu_0}\nabla\times\vec B-\vec j\eqno(1.3)$$
$$\eps_0\nabla\cdot\vec E=\rho\eqno(1.4)$$
and the equations in matter (1.1) and (1.2) plus
$${\d\vec D\over\d t}=\nabla\times\vec H-\vec j_c\eqno(1.5)$$
$$\nabla\cdot\vec D=\rho_c.\eqno(1.6)$$
Here we have used the standard notation with $\vec E$ and $\vec H$ being 
the electric and magnetic fields, $\vec B$ and $\vec D$ the magnetic induction
and electrical displacement and $\rho$, $\vec j$ the charge and current densities,
respectively. $\eps_0$ is the dielectric constant of the vacuum and
$\mu_0=1/(\eps_0c^2)$ where $c$ is the velocity of light.
Two new fields $\vec D$ and $\vec H$ appear in (1.5) and (1.6)
and $\rho_c$ and $\vec j_c$ are the charge and current densties,
respectively, in contrast to the microscopic charge and current
densities in (1.3) and (1.4). To have a closed system of equations
for dead or living matter additional so-called constitutive relations
are needed which take the structure of the matter into account.
The simplest such relations are
$$\vec B=\mu\vec H,\quad\vec D=\eps_0\vec E+\vec P\eqno(1.7)$$
where $\vec P$ is the electric polarization (see next section),
and Ohm's law
$$\vec j_c=\sigma\vec E.\eqno(1.8)$$

However, on the atomic scale matter is vacuum with charged nuclei
and electrons moving around. Consequently equations (1.1-4) are still
true in dead and living matter, but $\rho(t,\vec x)$, $\vec j(t,\vec x)$ 
vary rapidly in space and time. These rapid variations of the microscopic
quantities are not observed in general. Instead a spatial average
over many atoms or even many cells in the heart or brain, for example,
is measured. That means the macroscopic Maxwell's equations must be
$derived$ from the microscopic equations (1.1-4) by spatial
averaging. In this derivation the meaning of the conduction charge
and current densities $\rho_c$, $\vec j_c$ and their properties are found.
This point of view is shared by all modern authors of textbooks,
although it is sometimes a little hidden. For example Jackson in his
most-cited book [1] treats the macroscopic averaging in Sect.6.7.
But he considers the averaging of the charge density $\rho$ only,
leaving the more complicated current density $\vec j$ for ``those
readers who enjoy such challenges''. We are such readers, so we give the
full derivation in the next section. This is also necessary in order to
understand the properties of the different macroscopic currents. 
We here consider a living organ where the averaging must be performed 
over many cells. In this way we get a description of the electrical
activity of the organ from first principles, i.e. from Maxwell's theory.

On the other hand in physiology one often constructs models with quantities
which have no direct relation to Maxwell's theory. For example, in the
bidomain model [2] one introduces at each point in space {\it two} electrical
potentials $V_i$ and $V_e$ which refer to intra- and extracellular space.
This does not follow from Maxwell's theory: As we will see in the next
section the averaging of the microscopic Maxwell's equations gives {\it one}
averaged electric field $\langle \vec E\rangle$ plus corrections which are
represented by polarization $\vec P$ (see eq. (2.18) below). $\vec P$ has
different properties than the gradient of the membrane potential $V_i-V_e$
in the bidomain model (see after eq. (3.6)). The intra- and
extracellular currents that are introduced in the bidomain model are
meaningful on the {\it microscopic} scale if one considers processes in
single cells. As a consequence the model leads to some ``mesoscopic''
description of living tissue. Since we are interested in the measured
far-field potential we need a true {\it macroscopic} description.
The form of the {\it macroscopic} current 
and charge densities follows rigorously from the spatial averaging of 
the microscopic quantities (eqs. (2.16) (2.36) below) and so
cannot be freely assumed. The macroscopic description of living tissue
is given by the macroscopic Maxwell's equations, and this is what is
required for ECG etc. The bidomain model on the other hand may be useful
on some intermediate scale between single cell and macroscopic.

Another basic model in electrophysiology is the so-called volume
conductor which goes back at least to R. Plonsey [3] (see also [4] [5]). The 
argument is the following:  The total current density $\vec j$ is written
as the sum of the conduction current $\vec j_c$ plus a so-called
source current $\vec j_s$
$$\vec j=\vec j_c+\vec j_s.\eqno(1.9)$$
This starting point makes sense if $\vec j$ is the spatially
averaged microscopic current $\langle\vec j\rangle$ and $\vec j_s$
is the polarization current plus further corrections found in the
next section. Note that one is not free to add some ``source current''
to Maxwell's equations because the equations govern the whole
phenomenon, {\it including the active cells} (see next section).
 Then it is said that the displacement current
$\d\vec D/\d t$ can be neglected so that $\vec j$ in (1.9) is equal to
$\curl\vec H$ due to (1.5). But in equation (1.5) only the conduction current
$j_c$ appears, not $\langle\vec j\rangle$ !. This misunderstanding has
bad consequences: If the current is a curl, its divergence vanishes
$$\nabla\cdot\vec j=0=\nabla\cdot (\sigma\vec E)+\nabla\cdot\vec j_s
\eqno(1.10)$$
where Ohm's law (1.8) has been used. Assuming a quasi-static situation
where
$$\vec E=-\nabla V\eqno(1.11)$$
one obtains the electric potential $V$. For constant conductivity $\sigma$
one arrives at the Laplace equation
$$\nabla^2 V=\sigma\lap V=-\nabla\cdot\vec j_s.\eqno(1.12)$$

Here a physicist protests: a current density cannot be the source
of the electric potential. A static electric field must come from a
charge density, a current density generates a magnetic field. The solution
of (1.12) would be
$$V(\vec x)={1\over 4\pi\sigma}\int{(\nabla\cdot\vec j_s)(\vec y)\over
\vert\vec x-\vec y\vert}d^3y.\eqno(1.13)$$
Now one sees that it is misleading to use Ohm's law (1.8) to 
calculate the potential: for conductivity $\sigma=0$ the potential would
be infinite. But the whole argument would also apply to a completely isolating
dielectrics in a condenser where $V$ is always finite. Nobody knows which
conductivity $\sigma$ must be used in (1.13). If one says that $\nabla\cdot\vec j_s/
\sigma$ ``models'' the source of the signals, then one makes fictitious
physics and abandons Maxwell's electrodynamics as first principle.

The correct treatment is given in Sect.3. We must use Gauss' law (1.6) instead
of Ohm's law. The result is what everybody intuitively knows: the source of the
electric field in electrophysiology is the polarization charge density at the
membranes of the active cells. The conductivity plays no role here; if Ohmic currents
are present they would weaken the measured potentials. 
We also discuss the consequences of our findings for the problem of 
analyzing cardiac activity.

\section{Derivation of the macroscopic Maxwell's equations}

This derivation is of very general character. It is true for dead or living
matter and does not depend on the structural details of the matter.
To have a concrete picture we consider an organ (heart or brain for example)
consisting of cells and extracellular space. We divide the extracellular 
space into ``virtual cells'' of
approximately the same volume as the real cells, so that if we say ``cells''
we mean both types. The cells can move as for example in blood or ions
in extracellular space.

Macroscopic measurements in electrophysiology give mean values over a length scale
of, say $L\approx 10^{-3}$ m. Therefore, we average all microscopic quantities
over a volume $L^3$.
This averaging will be carried out by convolution with a positive
function $a(\vec x)$; we denote it by angular brackets
$$\langle\vec E(t,\vec x)\rangle\=d\int a(\vec x-\vec x')\vec E(t,
\vec x')\,d^3x'.\eqno(2.1)$$
The integral over $a(\vec x)$ must be equal to
$$\int a(\vec x)\,d^3x=1\eqno(2.2)$$
and otherwise we assume $a(\vec x)$ to be smooth with
a compact support of an extension of $L$, large compared with the dimension
of the cells. Such spatial averaging is the only way to derive true macroscopic
equations from microscopic ones.

Differentiating (2.1), we find
$${\d\over\d x_i}\langle\vec E(t,\vec x)\rangle=\int{\d\over\d x_i}
a(\vec x-\vec x')\vec E(t,\vec x')\,d^3x'$$
$$=\int a(\vec x-\vec x'){\d\over\d x'_i}\vec E(t,\vec x')\,d^3x'=
\Bigl\langle{\d\vec E(t,\vec x)\over\d x_i}\Bigl\rangle,\eqno(2.3)$$
 where we have shifted the derivative from $\vec x$ to $\vec x'$ and
performed an integration by parts. This means that averaging and spatial
derivatives can be interchanged. The same is trivially true for partial
derivatives with respect to $t$. Consequently, the homogeneous Maxwell's
equations (1.1) (1.2) are valid for the averaged fields, too,
$${\d\over\d t}\langle\vec B(t,\vec x)\rangle=-\nabla\times\langle\vec E(t,
\vec x)\rangle\eqno(2.4)$$
$$\nabla\cdot\langle\vec B\rangle=0.\eqno(2.5)$$
This was very cheap, but the inhomogeneous equations require much more
work. The reason is that the averaged densities $\langle\rho(t,\vec x) 
\rangle$ and $\langle\vec j(t,\vec x)\rangle$ can no longer be assumed
as given quantities as in microscopic electrodynamics. They usually depend 
on the local fields and on electrochemical processes, therefore, 
they must be worked out in detail,
taking the cellular structure of the matter into account.

We write the microscopic charge density as a sum
$$\rho(t,\vec x)=\sum_i\rho_i(t,\vec x),\eqno(2.6)$$
of the charge densities $\rho_i$ of the individual cells. {\it Charge and current
densities $\rho_i, \vec j_i$ of individual cells can change with time,
therefore, arbitrary physical and chemical processes in living cells
are included in the following derivation.}
The charge density $\rho_i$ of cell $i$ is
concentrated at its position $\vec y_i(t)$. In the average
$$\langle\rho(t,\vec x)\rangle=\sum_i\int a(\vec x-\vec x')\rho_i(t,\vec x')
\,d^3x'\eqno(2.7)$$
$a(\vec x-\vec y_i+\vec y_i-\vec x')$ is slowly varying over the
microscopic support of $\rho_i$. We therefore use a Taylor expansion
and neglect quadratic and higher contributions:
$$\langle\rho(t,\vec x)\rangle=\sum_i\int d^3x'\Bigl[a(\vec x-\vec y_i)
+\sum_k{\d a(\vec x-\vec y_i)\over\d x_k}(y_{ik}-x'_k)\Bigl]\rho_i(t,\vec x')
$$
$$=\sum_iq_ia(\vec x-\vec y_i)-\sum_i\vec p_i\cdot{\d a(\vec x-\vec y_i)
\over\d\vec x}.\eqno(2.8)$$
Here
$$q_i(t)=\int d^3x'\,\rho_i(t,\vec x')\eqno(2.9)$$
is the total charge of the cell $i$, and
$$\vec p_i(t)=\int d^3x'\,(\vec x'-\vec y_i(t))\rho_i(t,\vec x')
\eqno(2.10)$$
is its dipole moment with respect to the center $\vec y_i$. 
Writing the first term on the r.h.s. of (2.8) as follows
$$q_ia(\vec x-\vec y_i)=\langle q_i\delta(\vec x-\vec y_i)\rangle,
\eqno(2.11)$$
we see that it represents the cells by point charges (monopoles) at the
positions $\vec y_i$ of the cells. The sum
$$\sum_iq_ia(\vec x-\vec y_i)\=d\rho_c(t,\vec x)\eqno(2.12)$$
is the conduction charge density. Here only charged cells ($q_i\ne 0$) 
contribute, which explains the name ``conduction charge". The individual 
charges (ions and proteins) of neutral
cells are not resolved, they add up to zero in the integral (2.9).

The dipole term in (2.8) can be written as follows
$$\vec p_i\cdot{\d a(\vec x-\vec y_i)\over\d\vec x}=\nabla_x\cdot\vec p_i
a(\vec x-\vec y_i)=\nabla_x\cdot\langle\vec p_i\delta(\vec x-\vec y_i)\rangle.
\eqno(2.13)$$
This represents a dipole moment at the position of the cell. The cells are represented by their
lowest multipole moments. Summing over all cells, we get the so-called
polarization charge density
$$\sum_i\vec p_i\cdot{\d a(\vec x-\vec y_i)\over\d\vec x}\=d\nabla\cdot
\vec P(t,\vec x),\eqno(2.14)$$
where
$$\vec P=\sum_i\langle\vec p_i\delta(\vec x-\vec y_i)\rangle\eqno(2.15)$$
is the electric polarization. Omitting higher multipole contributions,
we obtain for the averaged charge density
$$\langle\rho\rangle=\rho_c-\nabla\cdot\vec P.\eqno(2.16)$$
Then averaging of Gauss' law (1.4) leads to
$$\eps_0\nabla\cdot\langle\vec E\rangle=\rho_c-\nabla\cdot\vec P.\eqno(2.17)$$
Introducing the new phenomenological field
$$\vec D=\eps_0\langle\vec E\rangle+\vec P,\eqno(2.18)$$
the macroscopic Gauss' law assumes the same form as the microscopic one
$$\nabla\cdot\vec D=\rho_c.\eqno(2.19)$$
The field $\vec D$ is usually called the electric displacement, because
the macroscopic electric field $\langle\vec E\rangle$ is displaced by
the polarization $\vec P$ in (2.18). The source of $\vec D$ is only
the conduction charge density $\rho_c$, not the total charge density (2.16).

In a similar manner we consider the current density
$$\vec j(t,\vec x)=\sum_i\vec j_i(t,\vec x),\eqno(2.20)$$
where $\vec j_i$ is the current density produced by the cell $i$. Charge
conservation for the cell $i$ implies
$$\int d^3x'\,\d_t\rho_i=-\int\nabla\cdot\vec j_i\,d^3x'=-\int\vec j_i\cdot
d\vec\sigma=0,\eqno(2.21)$$
where we have used Gauss' theorem and the vanishing of $\vec j_i$ at infinity.
Similarly we treat the next moments
$$\int d^3x'\,\d_t\rho_ix'_k=-\int d^3x'\,x'_k\nabla\cdot\vec j_i=$$ 
$$=-\int d^3x'\,\d'_l(x'_kj_{il})+\int d^3x'\,j_{ik}=\int d^3x'\,
j_{ik}\eqno(2.22)$$
$$\int d^3x'\,\d_t\rho_ix'_kx'_l=-\int d^3x'\,x'_kx'_l\nabla\cdot\vec j_i=$$
$$=\int d^3x'\,x'_lj_{ik}+\int d^3x'\,x'_kj_{il}.\eqno(2.23)$$

Now we are ready to expand the averaged current density
$$\langle\vec j_i\rangle=\int d^3x'\,\Bigl[a(\vec x-\vec y_i)+{\d a(
\vec x-\vec y_i)\over\d\vec x}\cdot(\vec y_i-\vec x')\Bigl]\vec
j(t,\vec x')\eqno(2.24)$$
$$\=d\vec A_i+\vec B_i,$$
By means of (2.22, 21), the first term $\vec A_i$ in (2.24) can be 
expressed by the dipole moment (2.10)
$$\vec A_i=a(\vec x-\vec y_i)\int d^3x'\,\d_t\rho_i(t,\vec x')(\vec x'
-\vec y_i(t))$$
$$=a(\vec x-\vec y_i)\d_t\vec p_i+a(\vec x-\vec y_i)q_i\d_t\vec y_i$$
$$=\langle(\d_t\vec p_i)\delta(\vec x-\vec y_i)\rangle+\langle q_i
(\d_t\vec y_i)\delta(\vec x-\vec y_i)\rangle.\eqno(2.25)$$
On the other hand, the time derivative of (2.15) yields
$$\d_t\vec P=\sum_i\langle(\d_t\vec p_i)\delta(\vec x-\vec y_i)
\rangle+\sum_i\langle\vec p_i\d_t\delta(\vec x-\vec y_i(t)\rangle$$
$$=\sum_i\langle(\d_t\vec p_i)\delta(\vec x-\vec y_i)\rangle-
{\d\over\d x_k}\sum_i\langle\vec p_i\delta(\vec x-\vec y_i)\d_t y_{ik}
\rangle.\eqno(2.26)$$
For the last term in (2.25) we introduce the conduction current
density
$$\vec j_c(t,\vec x)\=d\sum_i\langle q_i(\d_t\vec y_i)\delta(\vec x-
\vec y_i)\rangle.\eqno(2.27)$$
Then the sum of (2.25) over all cells yields
$$\sum_i\vec A_i=\d_t\vec P+{\d\over\d x_k}\sum_i\langle\vec p_i\delta(\vec x
-\vec y_i)\d_t y_{ik}\rangle+\vec j_c.\eqno(2.28)$$

To compute the second term $\vec B_i$ in (2.24), we decompose
$$\int d^3x'\,(y_{ik}-x'_k)j_{il}(t,\vec x')=$$
$$=\eh\int d^3x'\,\Bigl[(y_{ik}-x'_k)j_{il}-(y_{il}-x'_l)j_{ik}\Bigl]$$
$$+\eh\int d^3x'\,\Bigl[(y_{ik}-x'_k)j_{il}+(y_{il}-x_l')j_{ik}\Bigl]
\eqno(2.29)$$
into an antisymmetric and symmetric part. Using (2.22, 23) we get
$$=\eh\Bigl(\int d^3x'\,(\vec y_i-\vec x')\wedge\vec j_i\Bigl)_m+
\eh\Bigl[y_{ik}\int d^3x'\,\d_t\rho_ix'_l$$
$$-\int d^3x'\,x'_lj_{il}+y_{il}\int d^3x'\,\d_t\rho_ix'_k-\int
d^3x'\,x'_lj_{ik}\Bigl]=$$
$$=-{1\over\mu_0}(\vec m_i)_m+\eh\Bigl[y_{ik}\int d^3x'\,\d_t\rho_i
x'_l-\int d^3x'\,\d_t\rho_ix'_kx'_l+y_{il}\int d^3x'\,\d_t\rho_i
x'_k\Bigl].\eqno(2.30)$$
Here $\vec m_i$ is the magnetic moment of cell $i$ with respect
to its center $\vec y_i$, and the index $m$ is such that $k,l,m$ is a
cyclic permutation of 1,2,3. The last term in (2.30) leads to the
quadrupole moment 
$$q_{kl}^{(i)}=\int
d^3x'\,\rho_i(x'_k-y_{ik})(x'_l-y_{il}).\eqno(2.31)$$
Then (2.30) yields
$$=-{1\over\mu_0}(\vec m)_m-\eh\int d^3x'\,(\d_t\rho_i)(x'_k-y_{ik})(x'_l
-y_{il})$$
$$=-{1\over\mu_0}(\vec m_i)_m-\eh\Bigl[\d_tq_{kl}^{(i)}+\int d^3x'\,
\rho_i(\d_ty_{ik})(x'_l-y_{il})$$
$$+\int d^3x'\,\rho_i(x'_k-y_{ik})\d_ty_{il}\Bigl]=$$
$$=-{1\over\mu_0}(\vec m_i)_m-\eh\d_tq_{kl}^{(i)}-\eh[p_{il}\d_t
y_{ik}+p_{ik}\d_ty_{il}],\eqno(2.32)$$
where again the dipole moment (2.10) appears. According to (2.24),
this result must be multiplied by the gradient of $a(\vec x-\vec y_i)$
$$(\vec B_i)_l=-{1\over\mu_o}{\d a\over\d x_k}(\vec m_i)_m-{1\over 2}{\d a
\over\d x_k}(p_{ik}\d_ty_{il}+p_{il}\d_ty_{ik})$$
$$={1\over\mu_0}\Bigl(\nabla_xa(\vec x-\vec y_i)\wedge\vec m_i\Bigl)_l
-{1\over 2}{\d a\over\d x_k}(p_{ik}\d_ty_{il}+p_{il}\d_t y_{ik}).
\eqno(2.33)$$
This gives the following result for the vector $\vec B_i$
$$\vec B_i={1\over\mu_0}\nabla_x\times a(\vec x-\vec y_i)\vec m_i-{1\over 2}
{\d a\over\d x_k}(p_{ik}\d_t\vec y_i+\vec p_i\d_t
y_{ik}).\eqno(2.34)$$
Here we introduce the macroscopic magnetic moment density, or
magnetization
$$\vec M\=d\sum_ia(\vec x-\vec y_i)\vec m_i=\sum_i\langle\vec m_i
\delta(\vec x-\vec y_i)\rangle.\eqno(2.35)$$
Then the total contribution of $\vec B_i$ is equal to
$$\sum_i\vec B_i={1\over\mu_0}\nabla\times\vec M-{1\over 2}{\d\over\d x_k}\sum_i
\Bigl(\langle p_{ik}\delta(\vec x-\vec y_i)\d_t\vec y_i\rangle+$$
$$+\langle\vec p_i\delta(\vec x-\vec y_i)\d_t y_{ik}\rangle\Bigl).
$$
The last term herein can be combined with the middle term in (2.28).
We then obtain the following final result
$$\langle\vec j\rangle=\vec j_c+{\d\vec P\over\d t}+{1\over\mu_0}
\nabla\times\vec M+\vec V,\eqno(2.36)$$
with
$$V_l={1\over 2}{\d\over\d x_k}\Bigl\langle\sum_i(p_{il}\d_ty_{ik}
-p_{ik}\d_ty_{il})\delta(\vec x-\vec y_i)\Bigl\rangle.\eqno(2.37)$$

We notice that the divergence vanishes
$$\nabla\cdot\vec V={\d V_l\over\d x_l}={1\over 2}{\d^2\over\d x_l\d x_k}
\Bigl\langle\ldots\Bigl\rangle=0,$$
because the sum in (2.37) is antisymmetric in $k,l$, whereas the
second derivative is symmetric. Then the microscopic current conservation
$$\d_t\langle\rho\rangle+\nabla\cdot\langle\vec j\rangle=0$$
implies
$$\d_t\rho_c+\nabla\cdot\vec j_c=0\eqno(2.38)$$
by means of (2.16) and (2.36). Hence, the conduction current alone
is also conserved. For small velocities or small dipole moments, $\vec V$
(2.37) can be neglected, as well as the higher order multipole
contributions. Then the last macroscopic Maxwell's
equation, the macroscopic Amp\`ere's law assumes the following form
$$\eps_0{\d\langle\vec E\rangle\over\d t}={1\over\mu_0}\nabla\times\langle\vec
B\rangle-\vec j_c-{\d\vec P\over\d t}-{1\over\mu_0}\nabla\times\vec M.\eqno
(2.39)$$
Introducing the new field
$$\vec H={1\over\mu_0}\Bigl(\langle\vec B\rangle-\vec M\Bigl) 
\eqno(2.40)$$
and using the electric displacement (2.18), we finally get
$${\d\vec D\over\d t}=\nabla\times\vec H-\vec j_c.\eqno(2.41)$$
Unfortunately, the field $\vec H$ is usually called ``magnetic field" and
$\vec B$ magnetic induction, because it appears in the induction law
(2.4). Of course, $\vec B$ is the fundamental (microscopic) magnetic
field and $\vec H$ is derived from it (2.40). To avoid notational
confusion, we will simply say $\vec B$-field and $\vec H$-field in the
following. 

As mentioned above (see (2.16)) {\it this derivation of the macroscopic
Maxwell's equations is so general that it includes arbitrarily complicated
electrochemical processes in living cells.} Consequently, it is not possible
to introduce additional terms by hand for such processes. For example, a
''source current`` density (1.9) must be identified with some term in (2.36),
and one is not free in postulating properties of this quantity.

\section{Application to electrophysiology}

From now on all quantities are averaged macroscopic ones, so we omit the
averaging brackets.
It follows from the homogeneous Maxwell's equations (2.4) and (2.5) that the
$\vec E$ and $\vec B$ - fields can be expressed by the scalar and vector
potentials
$$\vec B=\nabla\times\vec A,\quad\vec E=-\nabla V-{\d\vec A\over\d t}.\eqno(3.1)$$
In the applications to electrophysiology we are concerned with time-dependent
electromagnetic fields in the frequency range 1-1000 Hz. The corresponding wave
lengths $\lambda$ are many kilometers. That means the distance of measurement
$r$ is always in the near zone $r\ll\lambda$. Then the magnetic field $\vec B$
can be neglected compared with the electric field $\vec E$, so we set $\vec A=0$
and we have the quasistatic approximation.
Now using $\vec E=-\nabla V$ in (2.19) we have
$$-\eps_0\nabla^2 V+\nabla\cdot\vec P=\rho_c\eqno(3.2)$$
or
$$\lap V=-{1\over\eps_0}(\rho_{\rm pol}+\rho_c)\eqno(3.3)$$
where
$$\rho_{\rm pol}=-\nabla\cdot\vec P\eqno(3.4)$$
is the polarization charge density.

It is no surprise that we have again obtained Laplace's equation for the electric
potential. However, the source of the potential on the right-hand side of (3.3) 
is completely different from $-\nabla\cdot\vec j_s/\sigma$ in (1.12). As already said the
appearance of the conductivity $\sigma$ in the denominator is misleading. Conduction
current cannot be at the origin of the electrical potential in a living organ. 
On the contrary, Ohm's
current is a dissipative process, it would diminish the potential.

The main source of the potential is the polarization-charge density $\rho_{\rm pol}$.
As we have seen in the last section it is different from zero if a cell $i$ has
an electric dipole moment $\vec p_i$ (2.10). Here is the point where living matter
differs from dead matter: In dead matter the polarization is generated by passive
response to an applied external electric field; in living matter an electric dipole
moment can be produced actively by chemical processes. For example, 
a dipole moment is generated, if $K^+$ ions move through the membrane out of a cell, 
while the negatively charged proteins remain inside [5], the exact details are
here not essential. The motion of the ions gives a conduction current which 
is not important for the electrical potential
because it does not enter on the right-hand side of (3.3). {\it What is important
is the separation of positive and negative charges which generates a dipole
moment and, hence, a polarization.}
If there remains a net charge in a cell, then it 
contributes to the charge density $\rho_c$. The total charge density 
$\rho_{\rm pol}+\rho_c$ is the source of the measured potential $V$.

In infinite space the solution of Laplace's equation (3.3) for a fixed time is given by
$$V(\vec x)={1\over 4\pi\eps_0}\int d^3y\,{(\rho_{\rm pol}+\rho_c)(\vec y)
\over \vert\vec x-\vec y\vert}$$
$$=-{1\over 4\pi\eps_0}\int d^3y\,{(\nabla\cdot\vec P)(\vec y)\over\vert\vec x-
\vec y\vert}+{1\over 4\pi\eps_0}\int d^3y\,{\rho_c(\vec y)\over\vert\vec x-\vec y\vert}.\eqno(3.5)$$
This solution can also be used in a nearly homogeneous situation as in the
heart chamber, then $\eps_0$ is the dielectric constant of the blood.
The first term in (3.5) is the polarization potential $V_{\rm pol}(\vec x)$ which can be 
identified with the total transmembrane activation potential of the organ. This is
the quantity of main clinical interest, so we concentrate on it in the following.

After partial integration we get
$$V_{\rm pol}(\vec x)={1\over 4\pi\eps_0}\int d^3y\,\vec P(\vec y)\cdot 
{\d\over\d\vec y}{1\over\vert\vec x-\vec y\vert}.\eqno(3.6)$$
Following Frank (1954) [7] and Dotti (1974) [8] we assume that the polarization
$\vec P\ne 0$ is located on a 2-dimensional surface $S$ at a certain time and
that the direction of $\vec P$ is normal to $S$; that means mathematically
$$\vec P(\vec x)=\vec n d(\vec x)\delta_S(\vec x),\eqno(3.7)$$
where $\delta_S(\vec x)$ is the delta-function on $S$. Such a surface of
active cell membranes is also called a dipolar wavefront; $d(\vec x)$ is the
dipole density. Then the corresponding polarization potential is equal to the
following surface integral over $S$
$$V_{\rm pol}(\vec x)={1\over 4\pi\eps_0}\int\limits_S d(\vec y){\d\over \d n}
{1\over\vert\vec x-\vec y\vert}\,d\sigma_y,\eqno(3.8)$$
where $\d/\d n$ is the derivative in the (outer) normal direction. The determination
of the dipole density $d(\vec x)$ from measured potential values 
$V_i=V_{\rm pol}(\vec x_i)$ is the so-called inverse problem.

In cardiology $S$ is taken to be the endocardium. The potential measurements
are performed either with a multielectrode balloon in the heart chamber or
non-invasively by body surface measurements. In the first case the best localization
of arrhythmia sources is achieved by contact mapping with a single electrode in contact
with the endocardium. To understand this let the point $\vec x$ of measurement
in (3.8) move towards a point $\vec x_S$ on the surface $S$. Then the limiting
potential value is given by [9]
$$V(\vec x_S)=-2\pi d(\vec x_S)+\int\limits_S d(\vec y){\cos\varphi_{xy}\over
\vert\vec x-\vec y\vert^2}\,d\sigma_y,\eqno(3.9)$$
where $\varphi_{xy}$ is the angle between the vector $\vec x-\vec y$ and the
normal $\vec n$. That means the measured potential $V(\vec x_S)$ directly 
gives the desired dipole density $d(\vec x_S)$ at the endocardium,
apart from the second term which may be considered as a background.

\end{document}